\documentclass[]{article}
\usepackage[utf8]{inputenc}
\usepackage{amsthm}
\usepackage{amsmath}
\usepackage{amssymb}
\usepackage{amsfonts}
\usepackage{algorithm}
\usepackage{algorithmicx}
\usepackage{authblk}
\usepackage{graphicx}
\usepackage{algpseudocode}
\usepackage{comment}
\usepackage{natbib}
\usepackage{hyperref}
\usepackage{xcolor}
\usepackage{setspace}
\usepackage{geometry}
\usepackage{bm}
 \usepackage{multirow}
\usepackage{subcaption}

\renewcommand{\det}{\textrm{det}}

\newcommand{\mvec}{\boldsymbol{m}}

\newcommand{\vvec}{\vec{\boldsymbol{v}}}
\newcommand{\xvec}{\boldsymbol{x}}
\newcommand{\yvec}{\boldsymbol{y}}
\newcommand{\zvec}{\boldsymbol{z}}

\newcommand{\Xmat}{\mathbf{X}}
\newcommand{\Zmat}{\mathbf{Z}}
\newcommand{\Pimat}{\boldsymbol{\Pi}}

\newcommand{\tauvec}{\boldsymbol{\tau}}

\algdef{SE}[DOWHILE]{Do}{doWhile}{\algorithmicdo}[1]{\algorithmicwhile\ #1}%

\usepackage{mathtools} 

\title{Mixture Modeling with Normalizing Flows for Spherical Density Estimation}
\author[1]{Tin Lok James Ng}
\author[2]{Andrew Zammit-Mangion}
\affil[1]{School of Computer Science and Statistics, Trinity College Dublin, Ireland}
\affil[2]{School of Mathematics and Applied Statistics, University of Wollongong, Australia}
\date{}

\begin{document}

\maketitle

\abstract
Normalizing flows are objects used for modeling complicated probability density functions, and have attracted considerable interest in recent years. Many flexible families of normalizing flows have been developed. However, the focus to date has largely been on normalizing flows on Euclidean domains; while normalizing flows have been developed for spherical and other non-Euclidean domains, these are generally less flexible than their Euclidean counterparts. To address this shortcoming, in this work we introduce a mixture-of-normalizing-flows model to construct complicated probability density functions on the sphere. This model provides a flexible alternative to existing parametric, semiparametric, and nonparametric, finite mixture models. Model estimation is performed using the expectation maximization algorithm and a variant thereof. The model is applied to simulated data, where the benefit over the conventional (single component) normalizing flow is verified. The model is then applied to two real-world data sets of events occurring on the surface of Earth; the first relating to earthquakes, and the second to terrorist activity. In both cases, we see that the mixture-of-normalizing-flows model yields a good representation of the density of event occurrence.

\section{Introduction}
Finite mixture models are widely used to model and analyze heterogeneous data. Of the several variants that appear in the literature, the parametric finite mixture models are the most popular; efficient estimation strategies to fit them to data are widely available, and their theoretical properties are well understood. It is common to model the mixture components using flexible density functions; for example both finite mixtures of skew normal densities as well as finite mixtures of $t$-densities have been used \citep{Fruhwirth2010, Lin2014, Hejblum2019}, while on the sphere both von Mises-Fisher densities and Kent densities have been used \citep{Banerjee2005, Peel2001, Gopal2014} as building blocks. The mixture component densities of these models are reasonably flexible, however they still rely on strong model assumptions that may not be satisfied in some practical settings. Semiparametric and nonparametric finite  mixture models have been proposed in order to relax the strong assumptions implicit to fully parametric mixture models. The semiparametric and nonparametric models replace one or more of the mixture components by a nonparametric density with possible constraints such as symmetry and log-concavity \citep{Chang2007, Hunter2007, Bordes2010, Levine2011}. The process of fitting semiparametric and nonparametric mixture models tends to be more computationally intensive than that of fitting parametric ones. 

An alternative, straightforward mechanism for modeling complicated probability distributions is the normalizing flow. Normalizing flows only require the specification of a simple ``base'' distribution (e.g., a standard normal or a uniform distribution) and a series of invertible and differentiable transformations \citep[see, e.g.,][]{Papamakarios2021}. The density of a sample can be evaluated by computing the product of the density of the transformed sample under the base distribution and the associated change in volume induced by the series of transformations. The latter term is the product of the absolute Jacobian determinants for each transformation. Many flexible families of normalizing flows on the Euclidean space have been developed \citep{Kobyzev2020}, some of which exhibit a ``universal property'' \citep{Huang2018, Jaini2019, Ng2023}, in the sense that they can be used to approximate a large class of density functions arbitrarily well.

Despite their popularity and efficacy, normalizing flows have a weakness: they require large and/or deep architectures to approximate complex target distributions with arbitrary precision \citep{Cornish2020}. To address this shortcoming, \cite{Izmailov2020} proposed to model the reference density function as a mixture of Gaussian density functions with unknown mean and covariance parameters, while \cite{Dinh2019} proposed a framework involving domain partitioning and locally invertible functions. In the latter approach, the transport map is no longer required to be fully invertible but only piecewise invertible, leading to greater flexibility. \cite{Ciobanu2021} took a different approach, and proposed using a mixture-of-normalizing-flows model, where each component of the mixture model is a density parameterized by a normalizing flow with its own parameters. \cite{Pires2020} proposed a variational mixture-of-normalizing-flows model, which combines the flexibility of normalizing flows with the ability to exploit class-membership structure.  Model fitting is done via optimization of a variational objective, for which the variational posterior over class membership latent variables is parameterized by a neural network.
While normalizing flows have been extensively studied and utilized on the Euclidean domain, there are several cases where the data should be treated and analyzed as elements of a non-Euclidean manifold. A popular example is directional statistics which involves the analysis of data on the unit sphere, and this is the case we focus on in this work. Spherical data arise in many application domains including gene expression analysis \citep{Banerjee2005}, protein bioinformatics \citep{Mardia2021}, and astronomy \citep{Jupp1995}. While normalizing flows on the sphere and more general manifolds have been considered \citep[e.g.,][]{Gemici2016, Rezende2020}, they are generally less flexible than their Euclidean counterparts. This is a direct consequence of the difficulties that arise when working on arbitrary manifolds with complex geometric structure, which in turn leads to both modeling and computational challenges. 

To address these limitations, in this work we propose adapting the mixtures-of-normalizing-flows models that have been developed for the Euclidean domain \citep{Pires2020, Ciobanu2021} for use on a widely used non-Euclidean manifold: the sphere. Specifically, we develop a mixture modeling framework where each mixture component is a spherical normalizing flow. Our spherical normalizing flows are constructed using exponential map flows \citep{Sei2013, Rezende2020, Ng2022}; however, the proposed framework can be adapted to any spherical normalizing flow. We give the requisite background on normalizing flows on Euclidean spaces and spheres in Section \ref{sec_background}. In Section \ref{sec_method} we then present the proposed mixture modeling framework along with the expectation-maximization (EM) algorithm (and a variant thereof) for model estimation, which we implement efficiently using mini-batch stochastic gradient descent. In this section we also briefly discuss approaches for mixture-components-order selection. We showcase the proposed methodology through a simulation study, density estimation of earthquake events, and density estimation of terrorist activity on the surface of Earth in Section \ref{sec_empirical}. Section~\ref{sec_conclusion} discusses potential future research directions and concludes.

\section{Background}
\label{sec_background}
Given two probability measures $\mu_0(\cdot)$ and $\mu_1(\cdot)$ defined on spaces ${\cal X}$ and ${\cal Z}$, respectively, a transport map $T: {\cal X} \rightarrow {\cal Z}$ is said to push forward $\mu_0(\cdot)$ to $\mu_1(\cdot)$ if, for any Borel subset $B \subset {\cal Z}$,
\begin{eqnarray}
\label{pushforward_measure}
 \mu_1(B) = \mu_0(T^{-1}(B)) ,
\end{eqnarray}
where the inverse $T^{-1}(\cdot)$ is set valued; specifically, $ T^{-1}(\zvec) = \{ \xvec \in {\cal X}: T(\xvec) = \zvec \} .$ For an injective transport map $T(\cdot)$, \eqref{pushforward_measure} can be re-formulated as $\mu_0(A) = \mu_1(T(A)),$  for any Borel subset $A \subset {\cal X}$. Suppose that the measures $\mu_0(\cdot), \mu_1(\cdot)$ are absolutely continuous with respect to the Lebesgue measure, with densities $f_0(\cdot) $ and $f_1(\cdot)$, respectively. If the map $T(\cdot)$ is bijective with a differentiable inverse $T^{-1}(\cdot)$, we obtain the familiar change-of-variables formula 
 \begin{equation}\label{eq:cov_formula}
f_0(\xvec) = f_1(T(\xvec))|\det(\nabla T(\xvec))|, \quad \xvec \in \mathcal{X},
\end{equation}
 which expresses a complicated probability density $f_0(\cdot)$ in terms of a simpler density $f_1(\cdot)$ and a transport map $T(\cdot)$. The reference density $f_1(\cdot)$ typically has no unknown parameter, and the multivariate standard normal density and the uniform density on a compact domain are common choices. \cite{Marzouk2016} discuss various strategies for parameterizing the transport map $T(\cdot)$, which has also been done using deep learning models \citep[e.g.,][]{Papamakarios2017, Kobyzev2020}. It has been proved that under mild conditions arbitrarily complex probability density functions can be well approximated by neural network based transport maps \citep{Huang2018, Ng2023}. Recent approaches introduce more flexibility  by defining $T(\cdot)$ as a composition of multiple transformations, that is, as $T(\cdot) \equiv T^{(K)} \circ \cdots \circ T^{(1)}(\cdot)$, where $T^{(k)}(\cdot)$ transforms $\zvec^{(k-1)}$ into $\zvec^{(k)}$, with $\zvec^{(0)} \equiv \xvec$ and $\zvec^{(K)} \equiv \zvec$. The composition of multiple transformations is called a  normalizing flow in the machine learning literature. Given two bijective maps $T^{(1)}(\cdot), T^{(2)}(\cdot)$ with differentiable inverses, their composition $T^{(2)} \circ T^{(1)}(\cdot)$ remains bijective with a differentiable inverse. The Jacobian determinant of the resulting composition remains computationally tractable since, by the chain rule,
$$ \mbox{det}(\nabla (T^{(2)} \circ T^{(1)}(\xvec))) = \mbox{det}(\nabla T^{(1)}(\xvec)) \mbox{det}(\nabla T^{(2)} (T^{(1)}(\xvec)) ), \quad \xvec \in \mathcal{X}.$$

  While normalizing flows have largely been studied and used in Euclidean spaces, data are often naturally described on Riemannian manifolds such as spheres and tori.  A number of normalizing flow techniques for Riemannian manifolds have been proposed; some of these, such as the technique proposed by \citet{Gemici2016}, involve projecting to the Euclidean space before projecting back to the original manifold. These approaches, however, lead to singularities if the manifold is not diffeomorphic to $\mathbb{R}^d$, as in the case of the sphere. 
  Projections can be avoided by constructing normalizing flows directly on the manifold of interest. \cite{Rezende2020} proposed constructing normalizing flows on spheres and tori using M{\"o}bius transformations and spherical splines. \cite{Mathieu2020} proposed constructing continuous normalizing flows for Riemannian manifolds by solving ordinary differential equations on manifolds. Here, we focus on the exponential map flow, which was first proposed by \cite{Sei2013}, then extended by \cite{Rezende2020}, and then adapted to a (spherical) spatial point process setting by \cite{Ng2022}.

  Let $\phi(\cdot)$ be a \emph{wrapping potential function} (see \cite{Sei2013} for a definition), and let $\exp_{\xvec}(\cdot)$ denote the \emph{exponential map}; then $\exp_{\xvec}(\nabla \phi(\xvec))$ is a valid exponential map flow for $\xvec \in \mathbb{S}^{d-1}$.   
  Let $p \in \mathbb{Z}^+$, $\beta_i >0, \mvec_i \in \mathbb{S}^{d-1}$, and $\eta_i > 0$ for $i = 1,\dots,p,$ such that $\sum_{i=1}^{p} \eta_i = 1$. The wrapping potential function we use in this work is given by \citep{Rezende2020, Ng2022}
\begin{eqnarray}
\label{radial_flow}
 \phi(\xvec) = \sum_{i=1}^{p} \frac{\eta_i}{\beta_i} e^{\beta_i (\cos d(\xvec, \mvec_i) - 1 ) }, \quad \xvec \in \mathbb{S}^{d-1},
\end{eqnarray}
where $\beta_i, \mvec_i$ and $\eta_i, i = 1,\dots,p,$ are model parameters that need to be estimated. For an intuitive description of exponential maps and their behavior on the sphere, see \cite{Ng2022}. A normalizing flow on the sphere can be constructed by stringing together several exponential map flows of the form $\exp_{\xvec}(\nabla \phi(\xvec))$ through composition.


We note that while we restrict ourselves to the sphere, \cite{Cohen2021} has generalized the exponential map flow to arbitrary Riemannian manifolds. While the parameterization of the normalizing flows in \cite{Cohen2021} leads to a universal property where arbitrary $c$-concave functions on compact manifolds can be approximated arbitrarily well, their construction leads to a piecewise smooth map which is not differentiable everywhere, and hence difficult to fit in practice.

\section{Methodology}
\label{sec_method}

\subsection{Mixture-of-normalizing-flows model for spherical data}

We consider the case where the target measure $\mu_0(\cdot)$ and reference measure $\mu_1(\cdot)$ admit densities with respect to the Lebesgue measure on $\mathbb{S}^{d-1}$, and model the target density as a mixture of $G$ component densities. Let $f_{0,g}(\cdot)$ be the $g$-th mixture component of the target density of interest, and $\tau_g$ the corresponding weight, where $\tau_g \ge 0, g = 1,\dots,G,$ and $\sum_{g=1}^{G} \tau_g = 1$. We model $f_0(\cdot)$ as
$$
f_0(\xvec) = \sum_{g=1}^{G} \tau_g f_{0,g}(\xvec),\quad \xvec \in \mathbb{S}^{d-1},
$$
where each $f_{0,g}(\cdot), g = 1,\dots,G,$ is constructed using a normalizing flow. For ease of exposition we consider the case where the normalizing flows have the same functional form, but different parameters; that is, the case where $f_{0,g}(\cdot) \equiv f(\cdot\, ; \Theta_g), g = 1,\dots, G,$ where $\Theta_g$ is the set of parameters corresponding to the $g$-th component. Let the reference measure for each component be the uniform density. From \eqref{eq:cov_formula} we then have that
$$
f_{0,g}(\xvec) \equiv f(\xvec ; \Theta_g) \propto |\det(\nabla T(\xvec; \Theta_g))|, \quad \xvec \in \mathbb{S}^{d-1},
$$
where for $\mathbb{S}^2$ the constant of proportionality is equal to $\frac{1}{4\pi}$, and where we have explicitly notated the dependence of the transport map $T(\cdot)$ on the component-specific parameters $\Theta_g$. Now, consider a set of $K$ transport maps, $T^{(1)}(\cdot\,; \Theta_{g,1}),\dots,T^{(K)}(\cdot\,; \Theta_{g,K})$ parameterized using parameters collected in $\Theta_g  \equiv \{\Theta_{g,1},\dots,\Theta_{g,K}\}$. We model the transport map for each mixture component as a composition of $K$ maps:
$$
T(\xvec; \Theta_g) = T^{(K)}\circ\cdots\circ T^{(1)}(\xvec; \Theta_g), \quad g = 1,\dots,G,
$$
where each $T^{(k)}(\cdot\,; \Theta_{g,k}) = \exp_{\xvec}(\nabla \phi(\xvec; \Theta_{g,k}))$ is an exponential map with wrapping potential function of the form \eqref{radial_flow}. Specifically, this wrapping potential function is given by
\begin{eqnarray}
\label{radial_flow2}
 \phi(\xvec; \Theta_{g,k}) = \sum_{i=1}^{p} \frac{\eta_i^{(g,k)}}{\beta_i^{(g,k)}} e^{\beta_i^{(g,k)} (\cos d(\xvec, \mvec_i^{(g,k)}) - 1 ) }, \quad \xvec \in \mathbb{S}^{d-1}.
\end{eqnarray}
Therefore, for $g=1,\ldots, G$, $\Theta_g \equiv  \{ \beta_i^{(g,k)}, \mvec_i^{(g,k)}, \eta_i^{(g,k)}: k=1, \ldots, K; i=1,\ldots,p \} $ are the model parameters for the $g$-th mixture component that together construct a composition of $K$ radial flows, where the $k$-th map in the composition has model parameters $\{ \beta_i^{(g,k)}, \mvec_i^{(g,k)}, \eta_i^{(g,k)}: i =1,\dots,p \}$.  Note that since we are assuming that each mixture component has the same functional form,  we are fixing the number of layers $K$ and the number of basis functions $p$ that are used for each mixture component; this choice is made for convenience and is not a model requirement. In our previous work \citep{Ng2022}, we showed that a small value of $p$ (i.e., $p=1$ or $p=2$) and a moderate to large value of $K$ (between 20 and 40) led to good performance in practice; we found that this was the case with our mixture-of-normalizing-flows model as well (see Section~\ref{sec_empirical}).

Let $\Theta \equiv \{\Theta_1,\dots,\Theta_G\}$ and $\tauvec \equiv (\tau_1, \ldots, \tau_G)'$. The problem of density estimation on the sphere using our mixture-of-normalizing-flows model reduces to the problem of estimating $\Theta$ and $\tauvec$ from data. We do this using the EM and related algorithms, that are often used when fitting mixture models. 

\subsection{Parameter estimation using the EM Algorithm}\label{sec:EM}

In this section we present both the standard EM algorithm, as well as an adaptation of it that is often referred to as the hard EM algorithm \citep[e.g.,][]{Samdani_2012}. Both algorithms could be used to fit the mixture-of-normalizing-flows model; however the hard EM algorithm is more computationally efficient and offers a solution to the problem of determining the order (i.e., the number of components) of the mixture.

\subsubsection{The standard (soft) EM algorithm}

Given $N$ observations $\Xmat \equiv (\xvec_1, \ldots, \xvec_N)$ where each $\xvec_j \in \mathbb{S}^{d-1}, j = 1,\dots,N$, the likelihood function for the unknown parameters is given by
\begin{eqnarray}
\label{likelihood}
 L(\Theta, \bm{\tau}; \Xmat ) = \prod_{j=1}^{N} \left(\sum_{g=1}^{G} \tau_g f(\xvec_j ; \Theta_g)\right). 
\end{eqnarray}
To facilitate estimation with the EM algorithm, we introduce the latent variables $\Zmat \equiv (\zvec_1, \ldots, \zvec_N)$ that denote the latent assignment of observations to mixture components, where $\zvec_j \equiv (z_{j,1}, \ldots, z_{j,G})'$ and $z_{j,g}=1$ if observation $j$ belongs to the $g$-th mixture component and $z_{j,g}=0$ otherwise \citep[see, e.g., ][Ch.~9]{Bishop_2006}. Then, the so-called complete-data likelihood function is given by
\begin{eqnarray}
\label{complete_data_likelihood}
L_c(\Theta, \bm{\tau};\Xmat, \Zmat ) = \prod_{j=1}^{N} \prod_{g=1}^{G} \big(  \tau_g f(\xvec_j; \Theta_g) \big)^{z_{j,g} }.
\end{eqnarray}
Taking logarithms of \eqref{complete_data_likelihood} we obtain the complete data log-likelihood function:
\begin{eqnarray}
\label{complete_data_log_likelihood}
\ell_c(\Theta, \bm{\tau}; \Xmat, \Zmat  ) \equiv \log L_c(\Theta, \bm{\tau}; \Xmat, \Zmat ) = \sum_{j=1}^{N} \sum_{g=1}^{G} z_{j,g} \big( \log \tau_g + \log f( \xvec_j ; \Theta_g) \big).  
\end{eqnarray}

The EM algorithm can now be readily applied to predict the latent mixture component assignments $\Zmat$ in the E-step, and to estimate the unknown model parameters $\Theta$ and $\bm{\tau}$ in the M-step.  At the $(t+1)$-th iteration, the E-step computes $\hat{\pi}_{j,g}^{(t+1)} \equiv \mathbb{P}(z_{j,g}=1 \mid \xvec_j, \hat{\Theta}^{(t)}, \hat{\bm{\tau}}^{(t)})$, that is, the probability of the latent assignment for each observation when conditioning on the data $\xvec_j$ and the estimates of the model parameters at the $t$-th iteration ($\hat\Theta^{(t)}$ and $\hat{\bm{\tau}}^{(t)}$):
\begin{eqnarray}
\label{e_step}
 \hat{\pi}_{j,g}^{(t+1)} = \frac{ \hat\tau_g^{(t)} f(\xvec_j; \hat\Theta_g^{(t)})}{ \sum_{h=1}^{G} \hat\tau_h^{(t)} f(\xvec_j; \hat\Theta_h^{(t)}) },\quad j = 1,\dots,N; g= 1,\dots,G .
\end{eqnarray}
We collect these latent probability parameters in the $N \times G$ matrix $\hat\Pimat^{(t+1)}$.

At the $(t+1)$-th iteration, the M-step maximizes the conditional expectation of the complete data log-likelihood in \eqref{complete_data_log_likelihood} with respect to the parameters $\Theta$ and $\tauvec$, where the expectation is taken with respect to the conditional distribution of $\Zmat$ given the observations $\Xmat$ and parameter estimates $\hat\Theta^{(t)}, \hat{\bm{\tau}}^{(t)}$. This expectation, which we denote by $Q(\{\Theta,\tauvec\}; \{\hat\Theta^{(t)}, \hat\tauvec^{(t)}\})$, is given by
\begin{align}
Q(\{\Theta,\tauvec\}; \{\hat\Theta^{(t)}, \hat\tauvec^{(t)}\}) &= \mathbb{E}\big( \ell_c(\Theta, \tauvec; \Xmat, \Zmat ) \mid  \Xmat, \hat\Theta^{(t)}, \hat\tauvec^{(t)} \big) \label{Qfun} \\
&= \sum_{j=1}^{N} \sum_{g=1}^{G} \hat{\pi}_{j,g}^{(t+1)} \big( \log \tau_g + \log f( \xvec_j ; \Theta_g) \big).   \label{cond_expectation_log_like}
\end{align}
 The update for $\tau_g$, obtained by maximizing \eqref{cond_expectation_log_like} with respect to $\tau_g$, can be derived analytically for $g = 1,\dots,G$:
\begin{eqnarray} 
\label{update_tau}
\hat{\tau}_g^{(t+1)} = \frac{\sum_{j=1}^{N} \hat{\pi}_{j,g}^{(t+1)}}{N}, \quad g=1,\ldots, G .
\end{eqnarray}
The optimization of the conditional expected complete data log-likelihood \eqref{cond_expectation_log_like} with respect to $\Theta$ does not result in a closed-form update rule for $\hat\Theta^{(t+1)}$, and therefore the optimization needs to be done numerically. As with $\tauvec$,  the parameters $\Theta$ for each mixture component can be updated separately. Specifically, we define
\begin{eqnarray}
\label{Q_g}
 Q_g(\Theta_g; \hat\Theta_g^{(t)}) \equiv \sum_{j=1}^{N} \hat{\pi}_{j,g}^{(t+1)} \log f(\xvec_j ; \Theta_g).\quad g=1, \ldots, G,
\end{eqnarray}
and optimize $Q_g(\Theta_g; \hat\Theta_g^{(t)})$ with respect to $\Theta_g, g = 1,\dots, G$. To facilitate this step we use mini-batch stochastic gradient descent (SGD) with automatic differentiation (AutoDiff) in \emph{PyTorch} \citep{Paszke2017}. Mini-batch SGD updates model parameters using small batches of data rather than the entire dataset. Hence, each step performs a descent based on an unbiased estimate of the gradient rather than the exact gradient; however it is more computationally efficient than exact gradient descent, and the introduced stochasticity helps to avoid local optima.  As in \citet{Ng2022}, we found that SGD works very well with this model. Once estimates for $\Theta_g, g = 1,\dots,G,$ are found, the E-step is repeated, followed by the M-step again, and so on until convergence. We summarise this EM algorithm, which we term the \emph{soft} EM algorithm to contrast it with the \emph{hard} EM algorithm that we discuss next, in Algorithm \ref{em_algo}.

\begin{algorithm}[t!]
  \caption{Soft (standard) EM Algorithm \newline \textbf{Input:} $G, \{\xvec_j\}_{j=1}^{N}$  \newline \textbf{Output:}  Estimated parameters $ \hat{\Theta}, \hat{\bm{\tau}}$, and the estimated component probabilities, $\hat\Pimat$}  
  \begin{algorithmic}
    \State \texttt{Initialise $ \{ \hat{\Theta}_g^{(0)}, \hat{\tau}_{g}^{(0)} \}_{g=1}^{G} $}\\
    \texttt{Set t = 0}
    \Do
    \State \texttt{E-Step}
         \For{\texttt{$j=1, \dots, N$}}
              \For{\texttt{$g=1, \dots, G$}}
                  \State Compute $\hat{\pi}^{(t+1)}_{j,g}$ according to \eqref{e_step}
              \EndFor
         \EndFor
         \\
         \State \texttt{M-Step}
          \For{\texttt{$g=1, \dots, G$}}
                 \State Compute $ \hat{\tau}^{(t+1)}_{g} $ according to \eqref{update_tau}
                 \State Find $\hat{\Theta}_{g}^{(t+1)}$ by optimizing  \eqref{Q_g} with respect to $\Theta_g$ using SGD with AutoDiff
                 \EndFor\\
           \State $t \leftarrow t + 1$      
           \doWhile{Not Converged}\\

           \State $\hat\Pimat \leftarrow \hat\Pimat^{(t)}$
           \State $\hat\tauvec \leftarrow \hat\tauvec^{(t)}$
           \State $\hat\Theta \leftarrow \hat\Theta^{(t)}$
           
   \end{algorithmic}
    \label{em_algo}
\end{algorithm}

\subsubsection{The hard EM algorithm}

 The hard EM algorithm, also known as the classification maximization algorithm, is a practical alternative to the standard (soft) EM algorithm. In the soft EM algorithm, the E-step is used to find the full conditional distribution of $z_{j,g}$, for $j = 1,\dots,N$ and $g = 1,\dots,G$. In the hard EM algorithm, this distribution is summarized as a Kronecker delta function centered at the mode of the true conditional distribution. That is, at the $(t+1)$-th iteration, the E-step approximates $\mathbb{P}(z_{j,g}=1 \mid \xvec_j, \hat\Theta^{(t)}, \hat{\bm{\tau}}^{(t)}) \approx \mathbb{I}(z_{j,g} = \hat{z}^{(t+1)}_{j,g})$, where
\begin{eqnarray}
\label{e_step_hard}
 \hat{z}^{(t+1)}_{j,g} = \begin{cases}
 1 \quad g = \underset{g'}{\mathrm{argmax}}\, \{\hat{\pi}^{(t+1)}_{j,g'}\}_{g'=1}^{G} \\
 0 \quad \mbox{otherwise,}
\end{cases}
\end{eqnarray}
for $j=1, \ldots, N$, and $g = 1,\dots,G,$ where $\hat{\pi}^{(t+1)}_{j, g}$ is defined in Equation \eqref{e_step}. We collect these estimated mixture component assignments in the $N \times G$ matrix $\hat\Zmat^{(t+1)}$. The M-step is done similarly to the standard EM algorithm, except that now expectations in \eqref{Qfun} are taken with respect to this degenerate conditional distribution rather than the full (true) conditional distribution. For the M-step, the update for $\hat{\bm{\tau}}$ becomes
\begin{eqnarray}
\label{update_tau_hard}
\hat{\tau}_g^{(t+1)} = \frac{N_g^{(t+1)}}{N}, \quad g = 1,\dots, G,
\end{eqnarray}
where $N_g^{(t+1)} \equiv |\{j: \hat{z}^{(t+1)}_{j,g} = 1\}|$ is the number of observations allocated to the $g$-th mixture component. The update $\hat{\Theta}^{(t+1)}$ can be done separately for each mixture component $g$ by optimizing the following objective function
\begin{eqnarray}
\label{Q_g_hard}
 Q_g(\Theta_g; \hat\Theta_g^{(t)}) = \sum_{\left\{j: \hat{z}^{(t+1)}_{j,g} =1\right\}} \log f(\xvec_j ; \Theta_g) 
\end{eqnarray}
with respect to $\Theta_g, g=1,\ldots, G$. Note how \eqref{Q_g_hard} involves a summation over $N_g^{(t+1)}$ rather than over $N$ points as in \eqref{Q_g}. This leads to a simpler optimization with SGD that in turn leads to improved computational efficiency (and, in practice, also more stable estimates).  Furthermore, the hard EM algorithm offers a simple solution to the problem of determining the number of mixture components; this is discussed in Section \ref{sec_order}. The hard EM algorithm is summarized in Algorithm \ref{em_algo_hard}. 

\begin{algorithm}[t!]
  \caption{Hard EM Algorithm \newline \textbf{Input:} $G, \{\xvec_j\}_{j=1}^{N}$,  \newline \textbf{Output:}  Estimated parameters $ \hat{\Theta}, \hat{\bm{\tau}}$, and the estimated mixture component assignments, $\hat\Zmat $}  
  \begin{algorithmic}
      \State \texttt{Initialise $ \{ \hat{\Theta}_g^{(0)}, \hat{\tau}_{g}^{(0)} \}_{g=1}^{G} $}\\
    \texttt{Set t = 0}
     \Do 
         \State \texttt{E-Step}
         \For{\texttt{$i=1, \dots, N$}}
              \For{\texttt{$g=1, \cdots, G$}}
              \State Compute $\hat{\pi}^{(t+1)}_{j,g}$ according to \eqref{e_step}
              \State Compute $\hat{z}^{(t+1)}_{j,g}$ according to \eqref{e_step_hard}
              \EndFor
         \EndFor
         \\
          \State \texttt{M-Step}
          \For{\texttt{$g=1, \dots, G$}}
                 \State Compute $ \hat{\tau}^{(t+1)}_{g} $ according to \eqref{update_tau_hard}
                 \State Find $\hat{\Theta}_{g}^{(t+1)}$ by optimizing  \eqref{Q_g_hard} with respect to $\Theta_g$ using SGD with AutoDiff
                 \EndFor\\
     \State $t \leftarrow t + 1$
     \doWhile{Not Converged}\\
           \State $\hat{\Zmat} \leftarrow \hat{\Zmat}^{(t)}$
          \State $\hat\tauvec \leftarrow \hat\tauvec^{(t)}$
           \State $\hat\Theta \leftarrow \hat\Theta^{(t)}$
   \end{algorithmic}
    \label{em_algo_hard}
\end{algorithm}


\subsection{Order Selection}\label{sec_order}
Having a strategy to determine the number of mixture components $G$ in a finite mixture model is important. Many frequentist and Bayesian approaches have been proposed to select the optimal $G$ for finite mixtures of parametric distributions; these include modified likelihood ratio tests \citep{Dacunha-Castelle1999, Gassiat2002}, bootstrapping \citep{McLachlan1987}, information criteria approaches \citep{Spiegelhalter2002, Drton2017}, classification-based information criteria approaches \citep{Biernacki2000}, and marginal-likelihood-based methods \citep{Chib1995, Green1995}. Determining the number of components for mixtures of normalizing flows is more challenging since in addition to choosing $G$, one also needs to choose $K$ (the number of layers for the normalizing flows) and $p$ (the number of basis functions constructing the wrapping potential functions). While one may try to adapt some of the existing order selection approaches mentioned above to this context to find, jointly, optimal values for $G$, $K$ and $p$, these new approaches would need to be investigated both theoretically and practically. This lies beyond the scope of this paper. A potential computational bottleneck we envision with several of these approaches is that they typically require comparisons across a large number of fitted models; this could be prohibitive with mixture-of-normalizing-flows models.

In this work we sideline the issue of tuning $K$, $p$, and especially $G$ as follows. First, we leverage the good results we obtained from empirical studies with a conventional (single component) normalizing flow on $\mathbb{S}^2$ with $p = 1$ and $K = 20$ to fix $p$ and $K$ to those values, respectively. Second, for $G$, we fit the mixture-of-normalizing-flows model with a large number of mixture components (we set $G= 10$ in our simulation studies and $ G= 20$ in our studies on real data) using the hard EM algorithm, and then remove mixture components that have no observations allocated to them. This approach to mixture modeling is commonly referred to as `mixture overfitting' \citep[e.g.,][]{Rousseau2011, Van2015} and takes advantage of a tendency of mixture models to not assign observations to superfluous mixture components. We find this approach works well with our mixture-of-normalizing-flows model in the empirical studies of Section~\ref{sec_empirical}.

\section{Empirical Studies}
\label{sec_empirical}

This section showcases the mixture-of-normalizing flows model on $\mathbb{S}^2$ for a variety of point patterns. Section~\ref{sec_simulation} is a simulation study using a known density function that illustrates the potential benefit of having a mixture of normalizing flows rather than a single component normalizing flow when modeling data on the sphere. Sections \ref{sec_earthquake} and \ref{sec_terrorism} then show how our model can fit complex densities on the surface of Earth well through the use of an earthquake dataset and a terrorism dataset, respectively.

\subsection{Simulation Study}
\label{sec_simulation}

In this section we conduct a simulation experiment to compare the mixture-of-normalizing-flows model to the standard, single component, normalizing flow model \citep{Rezende2020} on $\mathbb{S}^2$. We generate synthetic data by simulating random observations from a mixture of von Mises-Fisher (vMF) densities on $\mathbb{S}^{2}$:
$$ \sum_{j=1}^{J} \pi_j f_{\textrm{vMF}}(\cdot\,; \mu_j, \kappa_j) ,$$ 
where $J$ is the number of mixture components for the mixture of vMF densities, $\pi_j, j=1, \ldots, J,$ are the mixture weights, and where $f_{\textrm{vMF}}(\cdot\,; \mu, \kappa)$ is the density of a vMF distribution with mean direction $\mu$ and concentration parameter $\kappa$. The density $f_{\textrm{vMF}}(\cdot\,; \mu, \kappa)$ converges to the uniform distribution on $\mathbb{S}^2$ when $\kappa$ goes to 0, and becomes increasingly concentrated at $\mu$ as $\kappa$ becomes larger. We randomly draw the mean directions from the uniform distribution on $\mathbb{S}^{2}$, and randomly draw the concentration parameters from the exponential distribution with rate parameter $\lambda$. We consider four different simulation setups by varying the number of mixture components $J$ and the rate paramter $\lambda$; specifically, we consider the cases $\{J,\lambda\} = \{10, 10^{-2}\}$, $\{J,\lambda\} = \{10, 10^{-3}\}$, $\{J,\lambda\} = \{20, 10^{-2}\}$, and $\{J,\lambda\} = \{20, 10^{-3}\}$. We note that the expected value of $\kappa$ under the simulation is inversely proportional to the rate parameter $\lambda$.

We fit both the mixture-of-normalizing-flows model with $G = 10$ mixture components, and the standard (single component) normalizing flow model \citep{Rezende2020} to the simulated datasets.  For both models, we set the number of compositions to $K=20$ and let $p=1$ in the wrapping potential function \eqref{radial_flow2}. We fit the (single component) normalizing flow model using the ``committee of networks'' approach adopted by \cite{Ng2022} in the point process setting. This strategy involves training several models (in our case 50) with random initializations, and then averaging their outputs; this was necessary since the fit of a normalizing flow tends to be highly sensitive to the initial parameter settings. We fit the mixture-of-normalizing-flows model using the hard EM algorithm; a committee of networks was not needed with the mixture-of-normalizing-flows model, likely because of the relative ease with which it can fit complicated densities due to the increased model flexibility. It took approximately 30 seconds to train the (single component) normalizing flow model, and approximately 15 minutes to train the mixture-of-normalizing-flows model; this increase in computing time was expected given that the mixture-of-normalizing-flows model contains an order of magnitude more parameters to estimate than the (single component) normalizing flow model.

For each simulation setting (combination of $J$ and $\lambda$), we repeated the simulation and fitting procedure 20 times, and computed the average and empirical standard deviation of the $L^1$ distance between the true and the estimated density functions for both approaches. The results are shown in Table \ref{sim_results_table}: The mixture-of-normalizing-flows model consistently outperforms the (single component) normalizing flow model. Note that both the (single component) normalizing flow model and the mixture-of-normalizing-flows model perform better for the ``simpler'' density functions, corresponding to when $J$ is smaller and when $\lambda$ is larger (which in turn leads to smaller mixture concentration parameters). 

\begin{table}
\caption{Average and empirical standard deviation (in parentheses) of the 20 $L^{1}$ distances between the true density function and the estimated density function in the simulation experiment for each simulation setting.}
\begin{center}
\begin{tabular}{ c| c| c| c| c  }
 Model  & $J = 10, \lambda = 10^{-2}$ & $J = 10, \lambda = 10^{-3}$ & $J = 20, \lambda = 10^{-2}$ & $J = 20, \lambda = 10^{-3}$  \\ \hline
NF &  1.42 (0.04) & 1.44 (0.04) & 1.50 (0.05) & 1.67 (0.07) \\
Mix NF &0.60 (0.03)& 0.67 (0.04) & 0.88 (0.15)  & 0.94 (0.16) 
\end{tabular}
\end{center}
\label{sim_results_table}
\end{table}

\subsection{Earthquake locations}\label{sec_earthquake}
To showcase the flexibility of the mixture-of-normalizing-flows model we also apply it real data that exhibits a complex structure. In our first setting we consider the locations of 7354 known earthquake events with body-wave magnitude above 6.0 that occurred between the years 1960 to 2018 across the globe. These data were extracted from the Geocoded Disasters (GDIS) dataset \citep{Rosvold2021}.  For each mixture component, we set the number of compositions to $K=20$ and let $p=1$ in the wrapping potential function \eqref{radial_flow2}. We fit the mixture-of-normalizing-flows model to the data with $G=20$ mixture components with the hard EM algorithm. The hard EM algorithm identified 17 non-empty mixture components.

\begin{figure}[t!]
\begin{center}
\includegraphics[width=\linewidth]{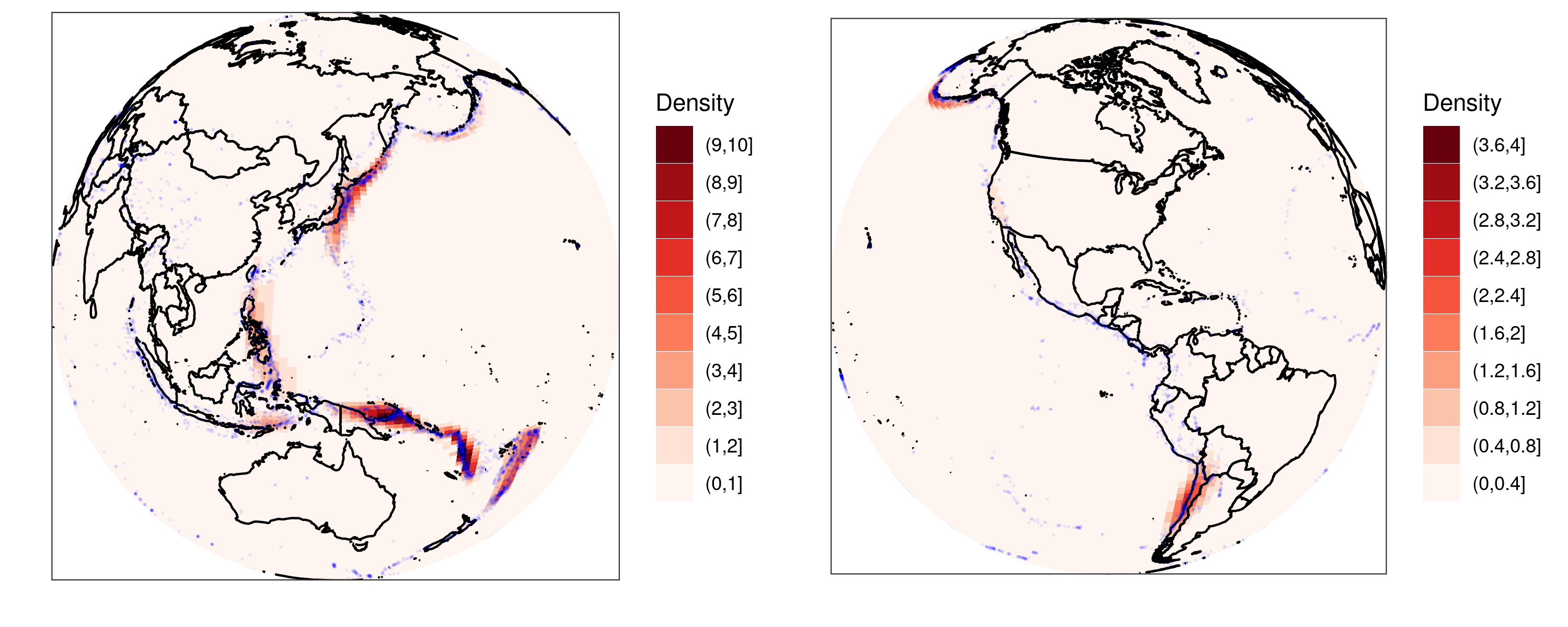}
\end{center}
\caption{Earthquake events (blue dots) and estimated density of earthquake locations obtained using the mixture-of-normalizing-flows model with $G = 20$ mixture components (red shading). (Left panel) View of Earth centered on 140$^\circ$E. (Right panel) View of Earth centered on 90$^\circ$W.} \label{earthquake_plot}
\end{figure}

\begin{figure}[t!]
\begin{center}
\includegraphics[width=\linewidth]{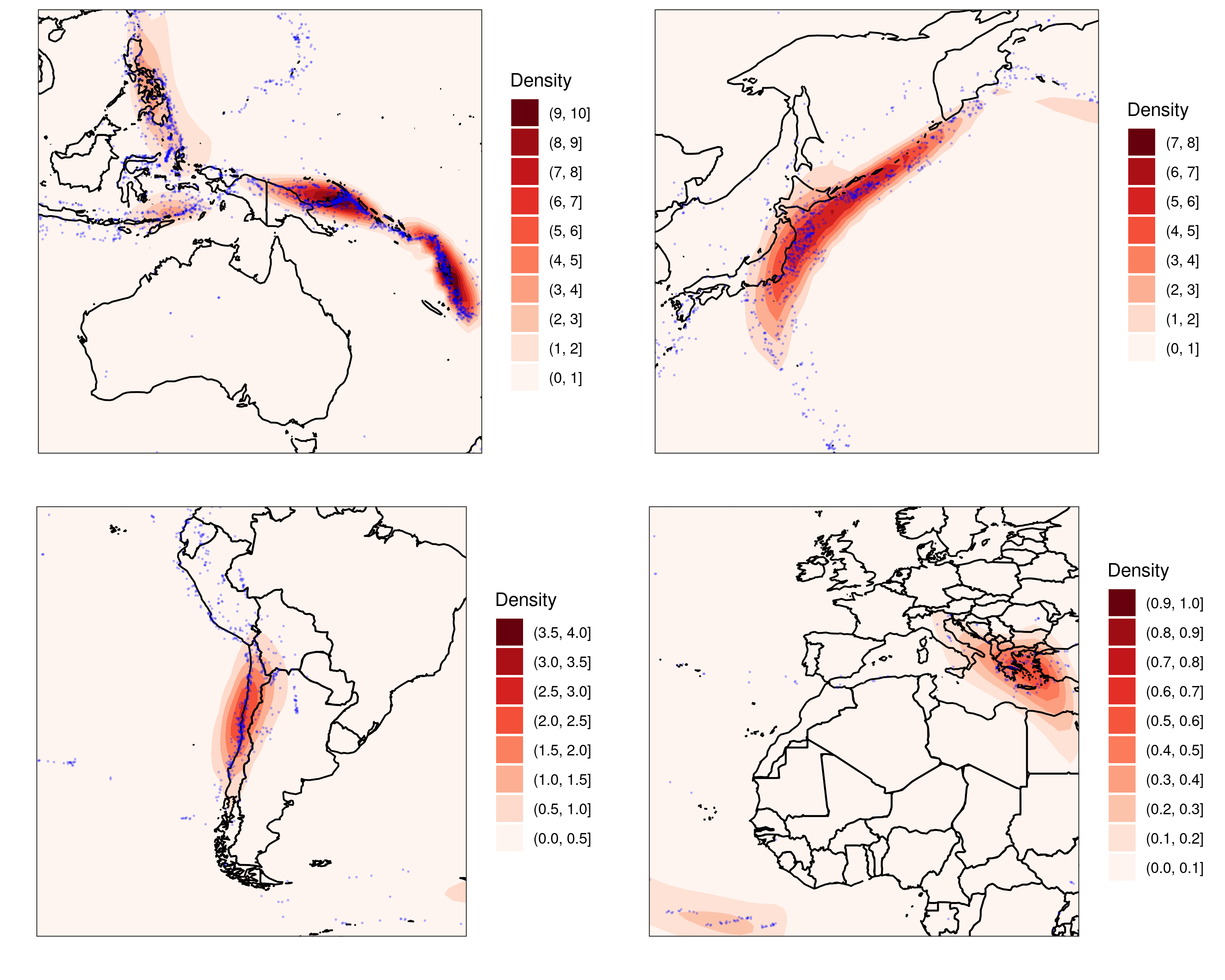}
\end{center}
\caption{Earthquakes between 1960 to 2018 (blue dots) and the corresponding estimated density obtained using the mixture-of-normalizing-flows model with $G = 20$ mixture components (red shading). (Top-left panel) Earthquakes in the region of the Pacific Islands, Papua New Guinea, Indonesia, and the Philippines. (Top-right panel) Earthquakes in the region of Japan and the Kuril Islands. (Bottom-left panel) Earthquakes in South America. (Bottom-right panel) Earthquakes in Europe and off the Liberian coast in the Atlantic Ocean.} \label{earthquake_zoomins}
\end{figure}

The density (relative to the uniform distribution on the unit sphere) using our model is shown in Figure~\ref{earthquake_plot}. The two orientations of Earth shown are chosen to depict the two most active regions on Earth: the western edge of the Pacific plate (left panel) and the western edge of the South American plate (right panel). Earthquakes largely occur where the tectonic plates meet, and thus predominantly follow long narrow paths along the surface of Earth; using a parametric model to fit the density of such data is extremely challenging (and would not be possible using most conventional parametric mixture models). Yet, we can see that the estimated density using the mixture-of-normalizing-flows model provides a very good fit to the data. We also show the fitted density at a selection of locations on Earth in Figure~\ref{earthquake_zoomins}; note how the mixture-of-normalizing-flows model is able to easily pick up multiple modes on the sphere of varying orientation, scale, and intensity.

We also attempted to fit a (single component) normalizing flows model fitted to the data using a ``committee of networks'' approach, where we average 50 models trained with random initializations. The fit we obtained, however, was inadequate and a poor representation of the data.

\subsection{Terrorism event locations}\label{sec_terrorism}
In our second setting we consider the locations of 8378 known terrorist events with known locations that occurred in the year 2020 across the globe. The data was extracted from the Global Terrorism Database (GTD).\footnote{\url{https://www.start.umd.edu/gtd/}} Like the earthquake dataset, these data are challenging to fit as terrorist activity tends to be highly clustered and largely influenced by geopolitical borders.  For each mixture component, we set the number of compositions to $K=20$ and let $p=1$ in the wrapping potential function \eqref{radial_flow2}. We fit the mixture-of-normalizing-flows model to the dataset with $G = 20$ mixture components using the hard EM algorithm. In this case, the hard EM algorithm identified 11 non-empty mixture components.

\begin{figure}[t!]
\begin{center}
\includegraphics[width=\linewidth]{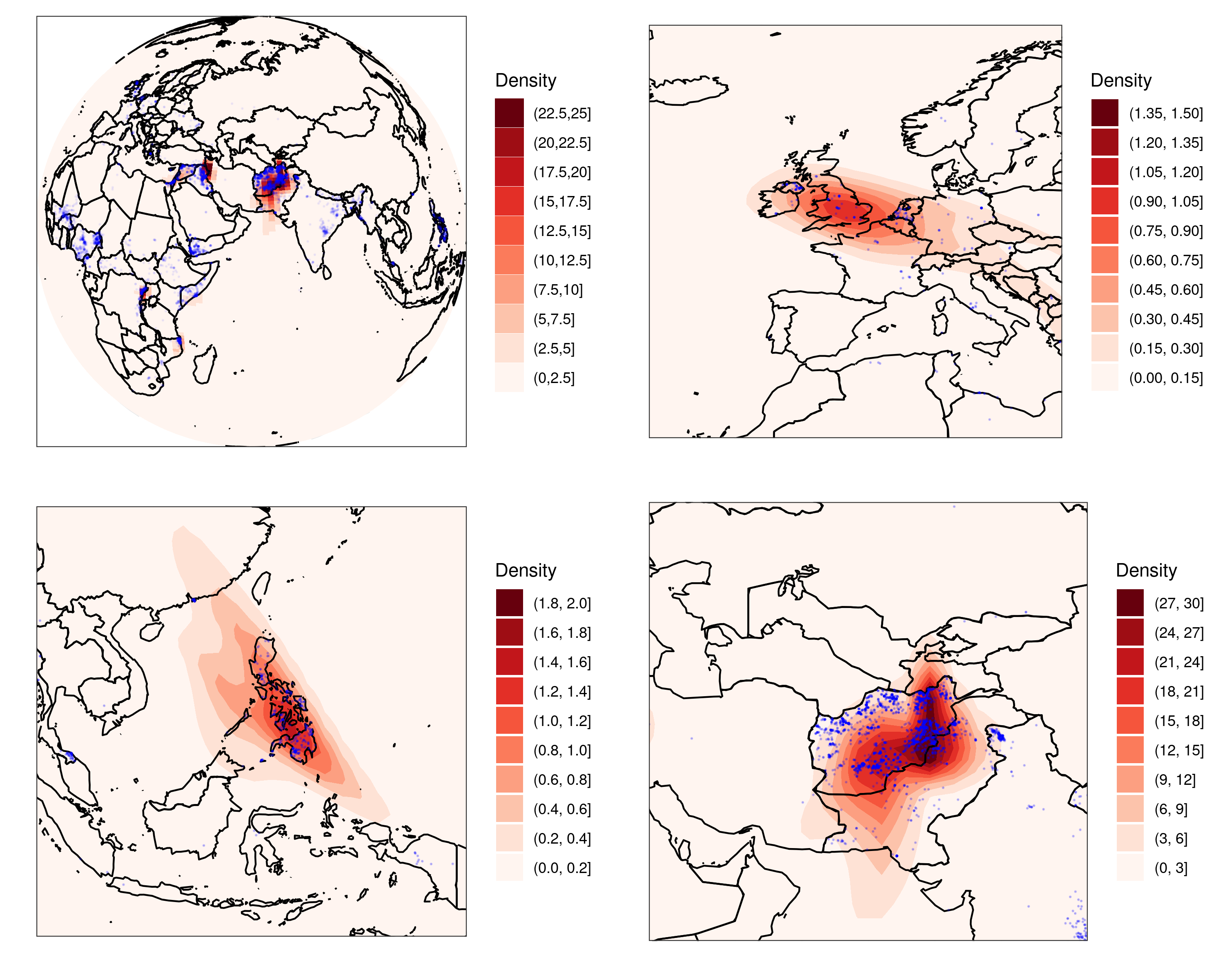}
\end{center}
\caption{Reported terrorist events in the year 2020 (blue dots) and the corresponding estimated density obtained using the mixture-of-normalizing-flows model with $G = 20$ mixture components (red shading). (Top-left panel) View of Earth centered on 60$^\circ$E. (Top-right panel) Events in Europe. (Bottom-left panel) Events in the Philippines. (Bottom-right panel) Events in Afghanistan.} \label{terr_plot}
\end{figure}

The density (relative to the uniform distribution on the unit sphere) using our model is shown in Figure~\ref{terr_plot}.  Similar to the earthquake dataset, we observe that the estimated density provides a good fit to the data.  The (single component) normalizing flow model was also fitted to the dataset using a ``committee of networks'' approach; here, as with the earthquake data, the normalizing flow  failed to achieve a reasonable fit to the data.

\section{Conclusion}
\label{sec_conclusion}
In this work we present a mixture modeling framework with normalizing flows for spherical density estimation. The proposed approach offers an attractive alternative to existing parametric and nonparametric mixture modeling approaches  by leveraging the computational efficiency and the representational power obtained when constructing deep hierarchies through function composition. The proposed EM algorithms are computationally efficient and scalable when used with mini-batch stochastic gradient descent.

An important consideration with mixture models that we have not explored is parameter identifiability. A necessary condition for identifiability is that there is a one-to-one map between the model parameters and the corresponding probability law (up to permutations of the mixture components). There is a long history of research on the identifiability conditions for finite mixtures of parametric distributions \citep{Teicher1961, Teicher1963, Barndorff-Nielsen1965, Holzmann2006}, and identifiability has been proved for various parametric families under various conditions. More recently, there has been an increased interest in the identifiability of nonparametric mixture models where each mixture component comes from a flexible, nonparametric family of probability distributions. Identifiability has been established for some of these models under various structural assumptions such as independence of marginal distributions and symmetry \citep{Hall2003, Hall2005, Hunter2007, Levine2011, DHaultfoeuille2015} and more general conditions \citep{Aragam2020}. Studying conditions under which identifiability holds for a mixture-of-normalizing-flows model is a challenging task. In order for the mixture-of-normalizing-flows model to be identifiable, each mixture component needs to be identifiable itself. Many popular normalizing flows models are typically parameterized by (deep) neural networks where the formulation of identifiability conditions is still in its early days \citep{Phuong2020, Bona-Pellissier2021}.

While the focus in this paper is on density estimation on spherical domains, the proposed methodology can be extended to more general manifolds. There are several open questions that warrant future research efforts. Identifiability, as discussed above, is one such open question. Another open question concerns interpretability: unlike parametric mixture models, the interpretation of the mixture components in a mixture-of-normalizing-flows model is challenging. Further, computationally-efficient ways for determining the optimal number of mixture components also needs to be addressed; this is challenging with any mixture model, but is particularly challenging with the mixture-of-normalizing-flows model where there are trade-offs between the number of mixture components and the complexity of each mixture component (via the number of compositions and the number of basis functions per wrapping potential function). Finally, model-based approaches for density estimation have the advantage that they lend themselves well to uncertainty quantification. With our mixture-of-normalizing-flows model one could, for example, employ the bootstrap \citep[e.g.,][]{Ng2022}. One could also develop Bayesian methods to recover posterior distributions over the model parameters as well as, potentially, the number of mixture components. 

\section*{Acknowledgements}

Andrew Zammit-Mangion’s research was supported by the Australian Research Council (ARC) Discovery Early Career Research Award DE180100203.

\bibliographystyle{asa}
\bibliography{manuscript}

\end{document}